\documentclass[preprint,review,12pt]{elsarticle}

\makeatletter
\def\@author#1{\g@addto@macro\elsauthors{\normalsize%
    \def\baselinestretch{1}%
    \upshape\authorsep#1\unskip\textsuperscript{%
      \ifx\@fnmark\@empty\else\unskip\sep\@fnmark\let\sep=,\fi
      \ifx\@corref\@empty\else\unskip\sep\@corref\let\sep=,\fi
      }%
    \def\authorsep{\unskip,\space}%
    \global\let\@fnmark\@empty
    \global\let\@corref\@empty  
    \global\let\sep\@empty}%
    \@eadauthor={#1}
}
\makeatother

\usepackage[left=3cm,top=2.5cm,right=3cm,bottom=2.5cm]{geometry} 
\usepackage{color}
\usepackage{verbatim}
\usepackage{dcolumn}  
\usepackage{bm}       
\usepackage{graphicx}
\usepackage{epstopdf}
\usepackage[english]{babel}
\usepackage{amsmath}
\usepackage{amssymb}
\usepackage{booktabs}
\usepackage{multirow}
\usepackage{hhline}

\biboptions{numbers,sort&compress}

\journal{Annals of Physics}

\begin{document}

\begin{frontmatter}



\title{Landau Levels in Uniaxially Strained Graphene: \\ A Geometrical Approach}

\author{Y. Betancur-Ocampo\corref{cor1}}
\ead{ybetancur@mda.cinvestav.mx}
\cortext[cor1]{corresponding author}
\author{M.E. Cifuentes-Quintal}
\author{G. Cordourier-Maruri}
\author{R. de Coss}
\address{Department of Applied Physics, Centro de Investigaci\'on y de Estudios Avanzados del Instituto Polit\'ecnico Nacional, A.P. 73 Cordemex 97310 M\'erida, Yucat\'an, M\'exico}

\begin{abstract}
The effect of strain on the Landau levels (LLs) spectra in graphene is studied, using an effective Dirac-like Hamiltonian which includes the distortion in the Dirac cones, anisotropy and spatial-dependence of the Fermi velocity induced by the lattice change through a renormalized linear momentum. We propose a geometrical approach to obtain the electron's wave-function and the LLs in graphene from the Sturm-Liouville theory, using the minimal substitution method. The coefficients of the renormalized linear momentum are fitted to the energy bands, which are obtained from a Density Functional Theory (DFT) calculation. In particular, we evaluate the case of Dirac cones with an ellipsoidal transversal section resulting from uniaxially strained graphene along the armchair (AC) and zig-zag (ZZ) directions. We found that uniaxial strain in graphene induces a contraction of the LLs spectra for both strain directions. Also, is evaluated the contribution of the tilting of Dirac cone axis resulting from the uniaxial deformations to the contraction of the LLs spectra. 

\end{abstract}

\begin{keyword}

Landau levels \sep Graphene \sep uniaxial strain \sep Dirac cones



\end{keyword}

\end{frontmatter}


\section{Introduction}
\label{intro}

Graphene is a two-dimensional material conformed by hexagonal rings of carbon atoms with unique physical properties, which make it a material of great scientific and technological interest. Graphene shows an anomalous Hall effect at room temperature, resulting from a linear dependence between the density
of electric charge carriers and the voltage \cite{f,e,Novo}. An extraordinary feature of the electronic properties in graphene is the linear relation between energy and momentum of $2p_z$ electrons at low energies (less than $1$ eV), being different than the usual quadratic energy momentum relations in ordinary materials. That dispersion relation ($E \propto k$) causes that the electron transport in graphene to be governed by a Dirac-like Hamiltonian, and electrons to behave like massless Dirac fermions \cite{e,Novo,DC3D,ImgLLs}. The peculiar electronic properties of graphene have consequences on the Landau Levels (LLs) spectra, which are defined as quantized energy states of charged particles in motion under a uniform applied magnetic field, $B$ \cite{b}. In an ordinary conductor, the energy of the LLs have a linear dependence with the quantization integer $n$, as $(n+1/2)B$. However, in graphene the LLs spectrum is quantized according to $\sqrt{nB}$ \cite{dre,Kopel,Zhang,l,k,Dietl,go,hat}.

In recent years, many works have been devoted to find practical methods to control graphene's properties. A possible way to modulate the electronic, vibrational and transport properties is by performing a deformation on the graphene sample \cite{tilt, c, Per09, Per10, Pellegrino, cho, ros, pele, juan, juan2, Bena, mon, Coc, gail, g, Ri, Naumov, h, Kitt, Crosse, i, Cad, Ho, Kumar, Jang, Shioya, car} or bilayer graphene \cite{much}. Thus, actual studies identify three main effects of uniaxial strain on low-energy band structure in graphene \cite{tilt, c, Per09, Per10, Pellegrino, cho}. First, a slipping of the Dirac points out of the high symmetry points K and K'. Second, a distortion in the traversal section of the Dirac cones, breaking the isotropy of the Fermi velocity; and third, a small vertical axis tilting of Dirac cones, which could be negligible in most of the situations. Considering that the LLs in graphene only depend on the shape of Dirac cone cross section, the second effect is expected to have more influence on the structure of LLs spectra. As we show here, the tilting of Dirac cone axis also has a non negligiable effect on LLs for high strains (15-20\%).

 
Some theoretical works use an anisotropic mass model to find the LLs in graphene through topological considerations \cite{hat}. Another authors frequently analyze the properties of strained graphene using the Tight-Binding (TB) approximation, including the effect of deformation on the atomic distances through the scaling of the hopping parameters \cite{go, tilt, c, Per09, Per10, Pellegrino, ros, pele, juan, juan2, Bena, mon, Coc, gail, g}. This hopping renormalization is commonly modeled with an exponential decay \cite{Pap} or using the Harrison's scaling rule $V_{pp\pi} \propto 1/l^2$ \cite{Harr}. Both renormalizations could fail beyond the linear elastic regime since the Poisson ratio changes with the strain. Indeed they could have a different dependence for each strain-type considered. An accurate and precise way to obtain the hopping parameters in strained graphene is through a fitting of the TB Hamiltonian to the energy bands obtained from Density Functional Theory (DFT) calculations \cite{Ri} or from experimental data. 

Instead of a hopping renormalization, an alternative way is proposed in the present work. We use a renormalized linear momentum in an effective Dirac-like Hamiltonian. The coefficients of the linear momentum are related with the geometrical parameters of the distorted Dirac cone, and can be calculated from a fitting to the energy bands obtained with a DFT calculation. Then, we apply a minimal substitution in the free-field effective Dirac-like Hamiltonian to get the LLs spectra. In particular, for uniaxial strain we found that the LLs spectra is contracted as a function of the deformation along the Zig-Zag (ZZ) and Arm-Chair (AC) directions. This contraction of the LLs spectra is due to the renormalization of the Fermi velocity with the strain, which is reduced by the stretching along these directions. In addition, we have evaluated the contribution of the tilting of Dirac cone axis to the contraction of the LLs in uniaxially deformed graphene 


\section{Landau levels in graphene}
\label{llg}

To clarify our methodology, we show how to obtain the LLs in unstrained pristine graphene from an effective Dirac-like Hamiltonian in a low energy and magnetic field regime. We treat the dynamics of an electron moving in a graphene sheet under a uniform magnetic field $\vec{B} = B\hat{z}$ perpendicular to the direction of propagation. We consider the electrons in graphene as massless Dirac fermions, having a linear dispersion relation $(E \propto k)$ \cite{Novo}. With $\vec{B} = 0$, the Hamiltonian $H$ only depends on the linear momentum $\vec{p}$ and can be represented by a 4 $\times$ 4 block diagonal matrix \cite{go,Bena}
\begin{equation}
H = v_F\left(\begin{array}{cc}
\vec{\sigma}\cdot\vec{p} & 0\\
0 & -\vec{\sigma}\cdot\vec{p}
\end{array}\right),
\label{hd}
\end{equation}

\noindent
where $v_F$ is the Fermi velocity and $\vec{\sigma}$ are the Pauli matrices acting on the pseudo-spin space, which discriminate between the contribution of the two triangular sublattices present in graphene. Each block in the Hamiltonian \eqref{hd} represents the $K$ and $K'$ valleys, coinciding with the high symmetry points in absence of strain. For $B \neq 0$, we do the minimal substitution $\vec{p} \rightarrow \vec{p} + e\vec{A}$ in the free-field Hamiltonian \eqref{hd} with the Landau gauge $\vec{A} = xB\hat{y}$ \cite{b,k}. Due to the $p_y$ conservation, the electron's wave-function in graphene can be expressed using variables separation, having the form
\begin{equation}
\vec{\Psi}(x,y) = \textrm{e}^{ik_yy}\vec{v}(x),
\label{sv}
\end{equation}

\noindent
where $\vec{\Psi}(x,y)$ and $\vec{v}(x) = (\vec{\phi}^{(+)}(x),\vec{\phi}^{(-)}(x))$ are four-component vector functions with $\vec{\phi}^{(+)}(x) = (f^{(+)}_A(x),g^{(+)}_B(x))$ and $\vec{\phi}^{(-)}(x) = (g^{(-)}_B(x),f^{(-)}_A(x))$, describing the pseudospin ($\{ f_A(x), g_B(x) \}$) and the valley ({\it K} or {\it K'}) states ($\{ \pm \}$). 

We focus in the $K$ valley, substituting \eqref{hd} and \eqref{sv} in the Dirac equation $H\vec{\Psi}(x,y) = E \vec{\Psi}(x,y)$, we obtain
\begin{equation}
\begin{array}{c}
v_F\left\{p_{ox} - ip_{oy}(k_y)\right\}g^{(+)}_B(x) = E f^{(+)}_A(x),\\
v_F\left\{p_{ox} + ip_{oy}(k_y)\right\}f^{(+)}_A(x) = E g^{(+)}_B(x),
\end{array}
\label{sed}
\end{equation}

\noindent
with $p_{ox} = p_x$ and $p_{oy}(k_y) = \hbar k_y + exB$. Decoupling the system of equations \eqref{sed}  and using the commutator $[p_x,p_{oy}(k_y)] = -i\hbar eB$, we get
\begin{equation}
v^2_F\{p^2_x + (\hbar k_y + e x B)^2 \} f^{(+)}_A(x) = \{E^2 - v^2_F \hbar eB\}f^{(+)}_A(x),
\end{equation}

\noindent
being similar to the quantum harmonic oscillator equation. Therefore, for Dirac electrons in the presence of a uniform perpendicular magnetic field, the LLs spectra is given by \cite{l,k}
\begin{equation}
E_n = \textrm{sgn}(n) \hbar \omega^D \sqrt{|n|}, \; \; \; \omega^D = v_F\sqrt{\frac{2eB}{\hbar}}, 
\label{elc}
\end{equation}

\noindent 
with $n = 0$, $\pm1$, $\pm2$, \ldots and degeneracy of $4SB/(h/e)$, where $S$ is the sample area and $(h/e)$ is the magnetic flux quantum. The four-fold degeneracy of LLs in deformed graphene $(n \neq 0)$ is due to a two-fold pseudospin and a two-fold valley degeneracy. While that for $n = 0$, there is a two-fold degeneracy because the valley index is the same as the sublattice index. This result is different from the LLs spectra for conventional conductors
$E_n = \hbar\omega(n + \frac{1}{2})$, where each level has a constant separation, while in graphene the LLs separation \eqref{elc} decreases as $|n|$
increases. This behaviour has been confirmed experimentally \cite{Novo, Zhang, l}. It is important to mention that fitting the expression \eqref{elc} to the experimental
spectrum, the value of $v_F$ can be obtained \cite{l}.

\section{Landau levels in strained graphene}
\label{llsg}



 It is known that the Dirac cones lose their isotropy when graphene is under non-isotropic strain and consequently, the effective Dirac Hamiltonian of \eqref{hd} is not valid anymore. To overcome this problem and thinking in a general case, i.e. an inhomogeneous strain \cite{pele, juan, juan2, Naumov, h, Kitt, Crosse, i}, we propose a renomalization in the linear momentum $\vec{p}$\, for the Hamiltonian \eqref{hd}
\begin{equation}
\tilde{\vec{p}} = a(x,y)p_x \hat{x} + b(x,y)p_y\hat{y},
\label{p'}
\end{equation}

\noindent
where $a(x,y)$ and $b(x,y)$ are dimensionless functions. In pristine graphene, the Dirac fermions move with an approximate $v_F$ of $10^6$ m/s \cite{e}. For strained graphene, the Fermi velocity is anisotropic and space-dependent \cite{pele,juan}. The present approach implicitly includes the anisotropy and spatial dependence of the Fermi velocity through the dimensionless functions $a(x,y)$ and $b(x,y)$. Substituting \eqref{p'} in the Hamiltonian \eqref{hd} for the $K$ valley 
\begin{equation}
H = v_F\left(\begin{array}{cc}
0 & p_xa(x,y) - ip_yb(x,y)\\
a(x,y)p_x + ib(x,y)p_y & 0
\end{array}\right),
\label{hd2}
\end{equation}

\noindent with a vector function $\vec{\Psi}(x,y) = (f^{(+)}_A, g^{(+)}_B)$, describing the pseudospin ($\{ f_A, g_B\}$). For a nonuniform magnetic or pseudomagnetic field $B_z$ perpendicular to the strained graphene sheet \cite{h, Kitt, Crosse}, we do the minimal substitution $\vec{p} \rightarrow \vec{\pi}^{\pm} = \vec{p} + e\vec{A}^{\pm}$ in the free-field Hamiltonian \eqref{hd2}, where $\vec{A}(x,y)$ is a potential vector that can depend of the inhomogeneous strain and acts different in to each valley \cite{h, Kitt, Crosse}. We focus in the $K$ valley, and decoupling the $2\times2$ equation system obtained from the Dirac equation $H\vec{\Psi} = E\vec{\Psi}$, we get
\begin{multline}
\Big\{(a\pi^+_x)^2 + (b\pi^+_y)^2 - \lambda_1(a\pi^+_x + ib\pi^+_y) + i[a\pi^+_x,b\pi^+_y]\Big\}f^{(+)}_A(x,y) = \frac{E^2}{v^2_F}f^{(+)}_A(x,y)\\
\Big\{(a\pi^+_x)^2 + (b\pi^+_y)^2 - \lambda_1(a\pi^+_x + ib\pi^+_y) - i[a\pi^+_x,b\pi^+_y] + \lambda_2\Big\}g^{(+)}_B(x,y) = \frac{E^2}{v^2_F}g^{(+)}_B(x,y)
\label{eg}
\end{multline}

\noindent where $(a\pi^+_x)^2 = (ap_x)^2 + 2ea^2A^{+2}_xp_x + e^2a^2A^{+2}_x - ie\hbar a\tfrac{\partial}{\partial x}(aA^+_x)$, a similar expression is obtained for $(b\pi^+_y)^2$ doing the changes $x \rightarrow y$ and $a \rightarrow b$. The commutator is given by $[a\pi^+_x,b\pi^+_y] = -i\hbar (a\tfrac{\partial b}{\partial x})(p_y + eA^+_y) + i\hbar (b\tfrac{\partial a}{\partial y})(p_x + eA^+_x) - i\hbar eabB^+_z$ with $B^+_z(x,y)$ the magnetic field generated by $\vec{A}^+$, the $\lambda_1(x,y) = i\hbar\tfrac{\partial a}{\partial x} + \hbar\tfrac{\partial b}{\partial y}$ and $\lambda_2(x,y) = \hbar^2\{-a\tfrac{\partial^2 a}{\partial x^2} - b\tfrac{\partial^2 b}{\partial y^2} + i(a\tfrac{\partial^2 b}{\partial x\partial y} - b\tfrac{\partial^2 a}{\partial x\partial y})\}$. For the $K'$ valley, the same decoupled equation system is obtained with $g^{(-)}_B(x,y)$ and $f^{(-)}_A(x,y)$ instead of $f^{(+)}_A(x,y)$ and $g^{(+)}_B(x,y)$ respectively, with superscripts $(-)$ in the $\hat{\pi}$ operators and $\vec{A}$ vector potential. The $2\times2$ partial differential equation system \eqref{eg} is an alternative way to describe the electron dynamics in strained graphene with deformed Dirac cones and in the presence of a nonuniform magnetic field from the anisotropy parameters with a spatial dependence. If we know the exact form of $a(x,y)$ and $b(x,y)$ from the dispersion relation or the Fermi velocity, it is possible to obtain the LLs solving the equation system \eqref{eg}. When an inhomogeneous unidirectional strain is applied, these functions must depend only on $x$ ($y$) due to the translational symmetry on $y$ ($x$). Assuming $x$ dependence for the anisotropy parameters and using the Landau gauge $\vec{A} = xB\hat{y}$, the $p_y$ component of the linear momentum is conserved. Thus, the equation system in \eqref{eg} is reduced to  
\begin{multline*}
-\frac{d}{dx}\left[a^2(x)\frac{d}{dx}f^{(+)}_A(x)\right] + \bigg\{\left(\frac{eBx}{\hbar} +  k_y\right)^2b^2(x) \\
+\frac{d}{dx}\left[a(x)b(x)\left(\frac{eBx}{\hbar}  + k_y\right)\right]\bigg\}f^{(+)}_A(x) = \frac{E^2}{\hbar^2v^2_F} f^{(+)}_A(x)
\end{multline*}
\begin{multline}
-\frac{d}{dx}\left[a^2(x)\frac{d}{dx}g^{(+)}_B(x)\right] + \bigg\{\left(\frac{eBx}{\hbar} +  k_y\right)^2b^2(x) \\
-\frac{d}{dx}\left[a(x)b(x)\left(\frac{eBx}{\hbar}  + k_y\right)\right] + R(x)\bigg\}g^{(+)}_B(x) = \frac{E^2}{\hbar^2v^2_F} g^{(+)}_B(x).
\label{raro}
\end{multline}


\noindent where $R(x) = 2b(x)\left(\tfrac{eBx}{\hbar}  + k_y\right)\tfrac{d a}{dx} - a(x)\tfrac{d^2 a}{dx^2}$. The equations \eqref{raro} are a general differential equations for the electron's wave function in nonuniform unidirectional strained graphene, which can be solved using the Sturm-Liouville theory \cite{SL} with eigenvalues $\lambda_n = E_n^2/\hbar^2v^2_F$.  

In previous studies on uniaxially strained graphene \cite{tilt, c, Per09, Per10, Pellegrino, ros, pele}, it was shown that the Dirac cones have an elliptical cross section for low energy regime $|E| < 1$ eV. As we shown in the Figure \ref{setcd}, the semi-major axis of the elliptical cross section is always along the tension direction. Hence, the functions $a(x)$ and $b(x)$ are approximately constants and they are related to the geometrical parameters of the ellipse, such as the semi-major axis ($A = E/a \hbar v_F$ for ZZ and $A = E/b \hbar v_F$ for AC direction) and semi-minor axis ($C = E/b \hbar v_F$ for ZZ and $C = E/a \hbar v_F$ for AC direction), or equivalently, scaling the energy axis with $q = E/\hbar v_F$, then $a = \cot \alpha = q/k_x$ and 
$b = \cot\beta = q/k_y$, where $\alpha$ and $\beta$ are the extremal elliptical cone angles. For pristine graphene $\alpha = \beta = \pi/4$. With $a$ and $b$ independent of $x$, the equation \eqref{raro} is reduced to the quantum harmonic oscillator equation. In the present case we obtain that the LLs spectra is given by

\begin{equation}
E_n = \textrm{sgn}(n)\sqrt{ab} \hbar \omega^D \sqrt{|n|}, \; \; \; \omega^D = v_F\sqrt{\frac{2eB}{\hbar}},
\label{ele}
\end{equation}

\noindent
with $n = 0$, $\pm1$, $\pm2$, \ldots and degeneracy of $4S'B/(h/e)$, where $S'$ is the deformed sample area. From equation \eqref{ele}, we can see that the strained graphene LLs depend on the geometrical parameters of the deformed Dirac cone, the respective cyclotron frequency $\omega^D$ and the quantum number 
$n$. Since the expressions \eqref{elc} and \eqref{ele} are similar, we have defined $\xi \equiv \sqrt{ab}$ as a parameter to measure the contraction ($\xi < 1$) or expansion ($\xi > 1$) of the LLs spectra under the same magnetic field. It should be noted that in others works, the concept of renormalized Fermi velocity $v^{*}_F$ is commonly used \cite{tilt}, and which is related with our parameter by $v_F^*=\xi v_F$. It is important to mention that when the cross section of the Dirac cone does not have an exact elliptical form, it is necessary to solve the differential equation \eqref{raro} to obtain the spectrum of LLs. In our best knowledge, there are not experimental reports on the effect of strain on the LLs spectra of graphene.


\section{Computational details}
\label{CD}

In order to know the geometrical parameters of strained graphene Dirac cones, DFT calculations were performed within the plane-wave pseudopotential framework, as implemented in the Quantum ESPRESSO (QE) package \cite{QE-2009}. The exchange-correlation functional was treated with the PBE parametrization of the generalized gradient approximation \cite{PBE}. Core electrons were replaced by an ultrasoft pseudopotential from the QE-PSLIB database \cite{QE-2009}, and valence wave functions (charge density) were expanded in plane waves with a kinetic energy cut-off of 40 Ry (320 Ry). 
\begin{figure*}[t!!]
     \begin{center}
     \resizebox{15.5cm}{!}{\includegraphics{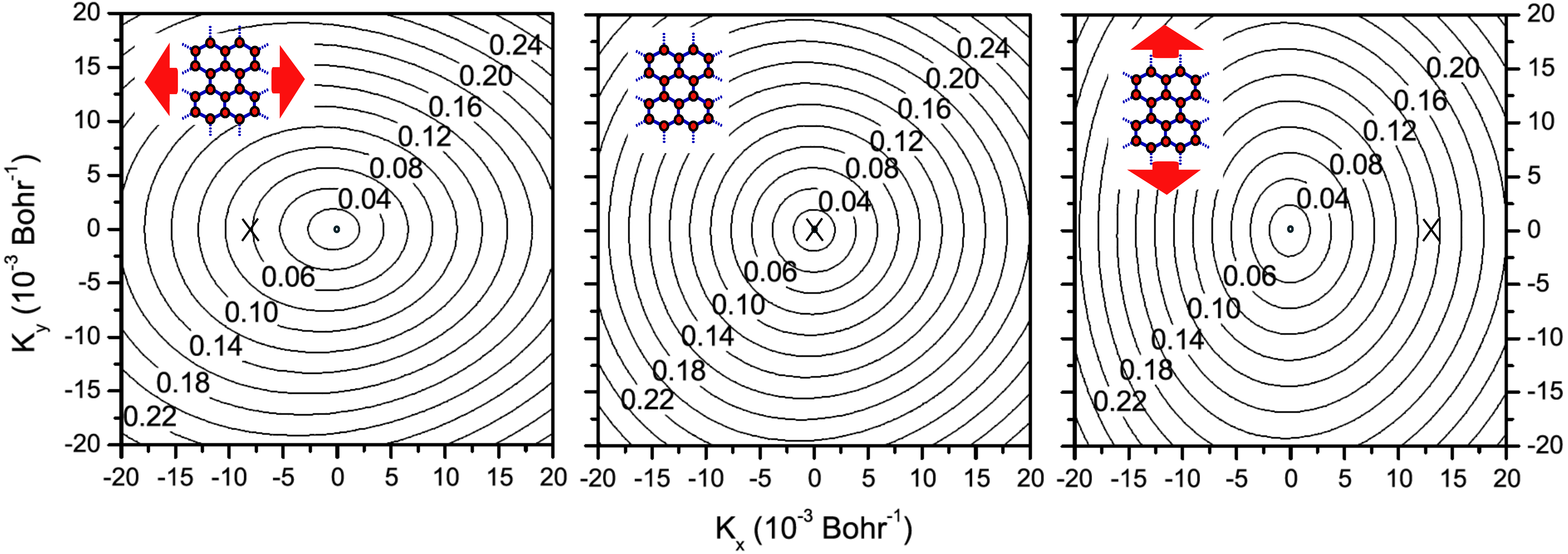}}
     \caption{Contour plots of the Dirac cones in pristine (center) and uniaxial strained (left and right) graphene. The uniaxial strain is along the ZZ (left) and AC (right) directions for a deformation of $10 \%$ for both cases. The origin of both axis corresponds to the Dirac point; note that the strain shifts the Dirac point from the K point. The cross marks refer the positions of the high symmetry K point in deformed graphene. The numbers on each curve indicates the value of the corresponding energy in eV.}
\label{setcd}
\end{center}
\end{figure*} 

We considered uniaxial strain up to 20\% of deformation along the AC and ZZ directions (see onsets in Figure \ref{setcd}). Because such amount of strain is beyond of the elastic regime of graphene, for a given strain, the Poisson ratio was obtained by a direct minimization of the electronic total energy. In each one of the steps to compute the Poisson ratio, we relax the carbons positions until the interatomic forces were 0.0001 Ry/Bohr or less.

For structural and energetic calculations, we employ a grid of 36$\times$36$\times$1 $k$-points within a 0.01 Ry of cold smearing \cite{mv-smearing}. The $E(k_x,k_y)$ surface was interpolated from a denser grid of 144$\times$144$\times$1 $k$-points and 0.001 Ry of smearing. In all cases, we left 10 \AA \, of vacuum space between successive layers to avoid spurious supercell interactions.

As expected, the Dirac cone cross section has an approximately elliptical shape, as shown Figure \ref{setcd}. We notice that the slipping of the Dirac points out of the high symmetry points K and K' becomes evident for a deformation of $10 \%$, for both deformation directions. In order to obtain the geometrical parameters of the deformed Dirac cones, we have fitted the conduction band around a Dirac point with an energy-cutoff of 0.3 eV for a range of deformation up to 20$\%$, using the dispersion relation $E = \hbar v_F\sqrt{a^2k^2_x + b^2k^2_y}$, where $a$ and $b$ are the fitting parameters and $(k_x,k_y)$ the linear momentum coordinates around the Dirac point. In our elliptical cone approximation, the values for the standard deviation are less than 0.003 eV, proving the accuracy of this approximation. 

\section{Results and Discussion}
\label{re}
\begin{figure}[t!!]
\centering
\begin{tabular}{cc}
(a) \qquad \qquad \qquad \qquad \qquad \qquad \qquad \qquad& (b) \qquad \qquad \qquad \qquad \qquad \qquad \qquad \qquad\\
\includegraphics[trim = 14mm 0mm 18mm 0mm, scale= 0.79, clip]{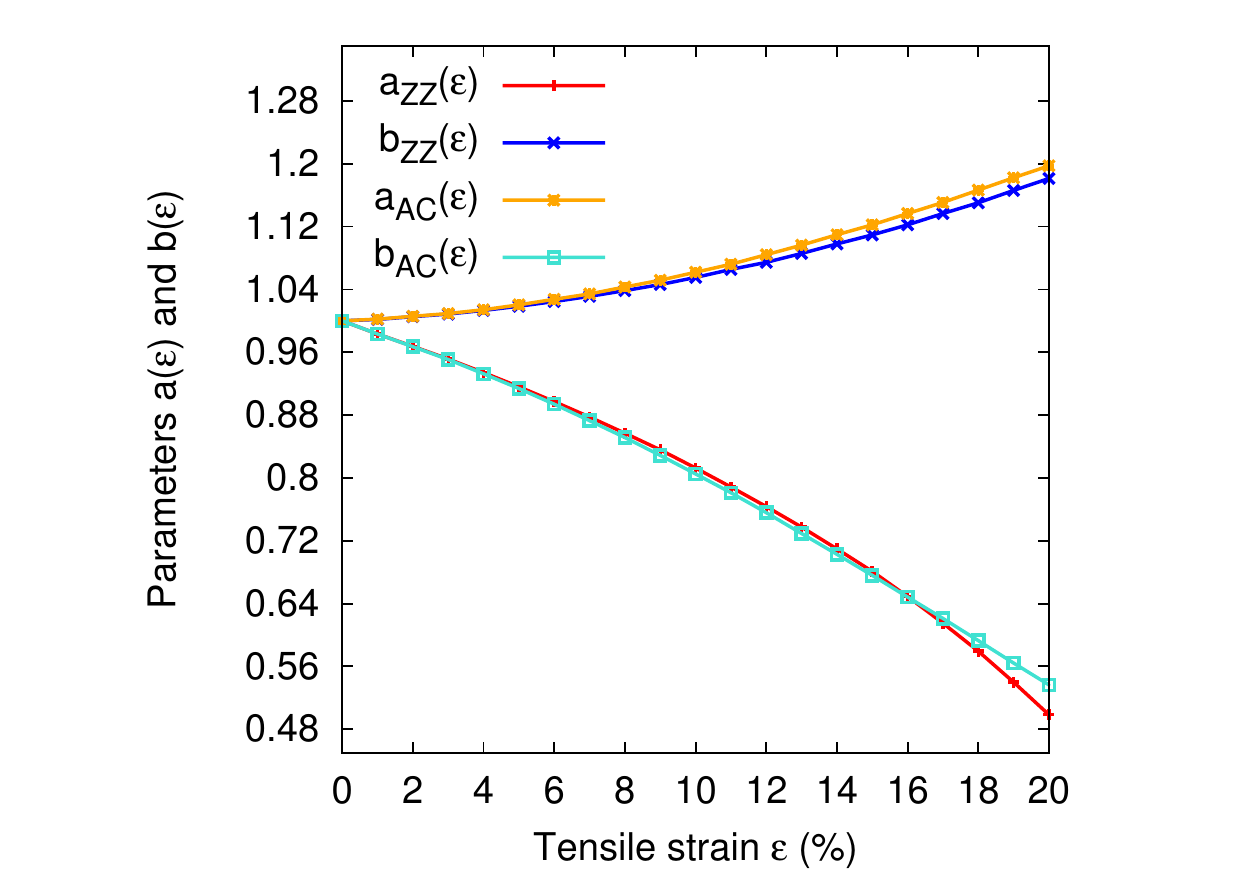} &
\includegraphics[trim = 14mm 0mm 18mm 0mm, scale= 0.79, clip]{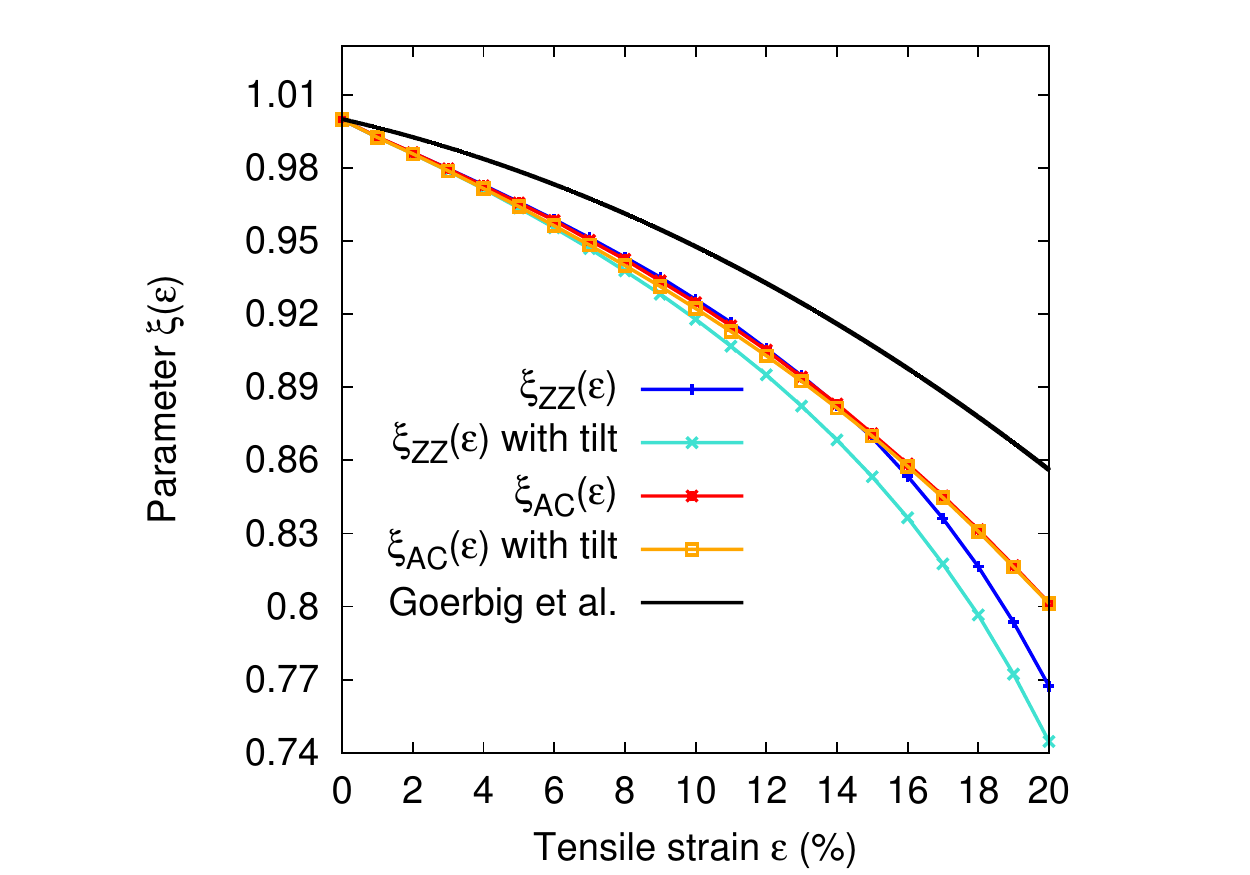}
\end{tabular}
\caption{{\small (a) Geometrical parameters $a$ and $b$ of the Dirac cone as a function of tensile strain ($\epsilon$) along the AC and ZZ directions. (b) Evolution of the $\xi$ parameter for strained graphene as a function of  tensile strain ($\epsilon$) along the AC (red line) and ZZ (blue line) directions with and without tilting of the Dirac cone axis. The black curve correspond to the renormalized Fermi velocity obtained from a TB approximation considering interactions up to second nearest neighbors in deformed graphene by Goerbig \emph{et al.} \cite{tilt}.}}
\label{as_bs}
\end{figure}

We found that for both strain directions (AC and ZZ), in the whole range of studied deformations, the LLs spectra is contracted ($\xi < 1$) with respect to the pristine case ($\xi = 1$) under the same magnetic field. It is interesting to note that the value of $\xi$, and hence the distance between LLs, decrease as the uniaxial deformation is increased. This behaviour can be explained in terms of the cyclotron orbit motion change induced by strain. Thus, if graphene is stretched, the cyclotron orbit motion has a mean-radius bigger than the equilibrium case, then the value of total energy decreases. From this perspective, the expansion case in the LLs spectra may occur when graphene sample is contracted.

In Figure \ref{as_bs} we show the plots of $a$, $b$ and $\xi$ for different values of deformation ($\epsilon$) along the AC and ZZ directions. From the Figure \ref{as_bs}(a) we observe that for $\epsilon < 10\%$, $a$ ($b$) in ZZ (AC) has approximately the same values than $b$ ($a$) in AC (ZZ). Thus, for a fixed
value of deformation, the fitted cones for AC and ZZ are practically the same but one rotated with respect to the other by $90^{\circ}$. Therefore, the effect of AC and ZZ uniaxial strains on LLs is expected to be practically the same for strain up to 10\% as we can see in the Figure \ref{as_bs}(b). This point can be explained if we consider that the parameter $\xi$ represents the ratio between the renormalized Fermi velocity and the Fermi velocity in pristine graphene. The renormalized Fermi velocity can be seen as an effective velocity of the anisotropic Fermi velocity in the cyclotron
motion. Considering that the cones have identical shape for AC and ZZ directions up to 10\%, we expect similar values of the renormalized Fermi velocity and similar contractions in the LLs spectra for both directions. Also, we observe that $\xi$ has a nearly linear behaviour in this range of deformations, with $\xi \approx 1 - 0.7\epsilon$. This result suggest that the degree of deformation ($\epsilon$) in a graphene sample can be estimated by extracting the value of $\xi$ from the LLs spectra. 

To evaluate the contribution of the tilting of the Dirac cone axis to the contraction of the LLs in uniaxially deformed graphene, we have fitted the conduction band using the dispersion relation of the form $E/\hbar v_F = a_ok_x + b_ok_y + \sqrt{a^2k_x^2 + b^2k_y^2}$, which is obtained diagonalizing the Weyl Hamiltonian \cite{go},  where $a_o$ and $b_o$ are responsible of the tilting of the Dirac cone axis. With an appropriate rotation of our system, we can use the expression $\xi = \sqrt{ab}(1 - a^2_o/a^2 - b^2_o/b^2)^{3/4}$ \cite{tilt}, instead of $\xi = \sqrt{ab}$. In all cases we obtain $b_o \approx 0$, having a tilting only in the $k_x$-axis, in agreement with others authors \cite{tilt}. The tilt can be neglected for strain up to 15\% as we can see in the Figure \ref{as_bs}(b), where the calculations of $\xi$ with and without tilting are shown. For strains along the ZZ direction, the tilt could have observable effects on the LLs spectra for deformations larger than 15\%. In order to compare the present results for $\xi$ with previous calculations reported in the literature, in Figure \ref{as_bs}(b) we include the result obtained from a TB effective model reported in \cite{tilt}, where a Harrison's scaling rule for the hopping parameters was used. We can see that the TB approximation, predicts a descreasing behaviour of the renormalized Fermi velocity as a function of strain, in qualitative agreement with the DFT calculations, but there is an important quantitative difference, which can be attributed to the relaxation of the carbons positions and the Poisson ratio changes beyond of the elastic regime included in the DFT calculations. 
\begin{figure*}[h!!]
\centering
\begin{tabular}{cc}
(a) \qquad \qquad \qquad \qquad \qquad \qquad \qquad \qquad& (b) \qquad \qquad \qquad \qquad \qquad \qquad \qquad \qquad\\
\includegraphics[trim = 14mm 0mm 18mm 0mm, scale= 0.79, clip]{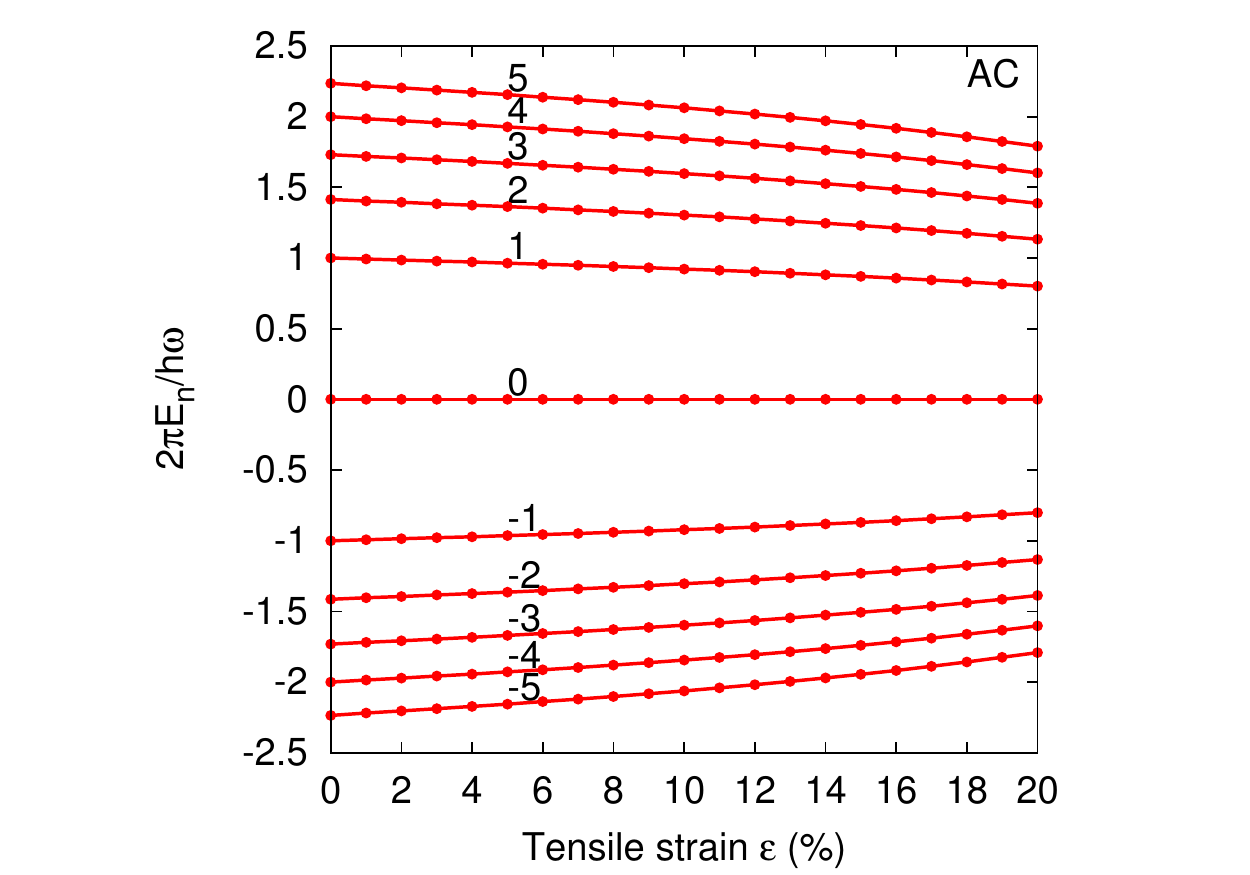} &
\includegraphics[trim = 14mm 0mm 18mm 0mm, scale= 0.79, clip]{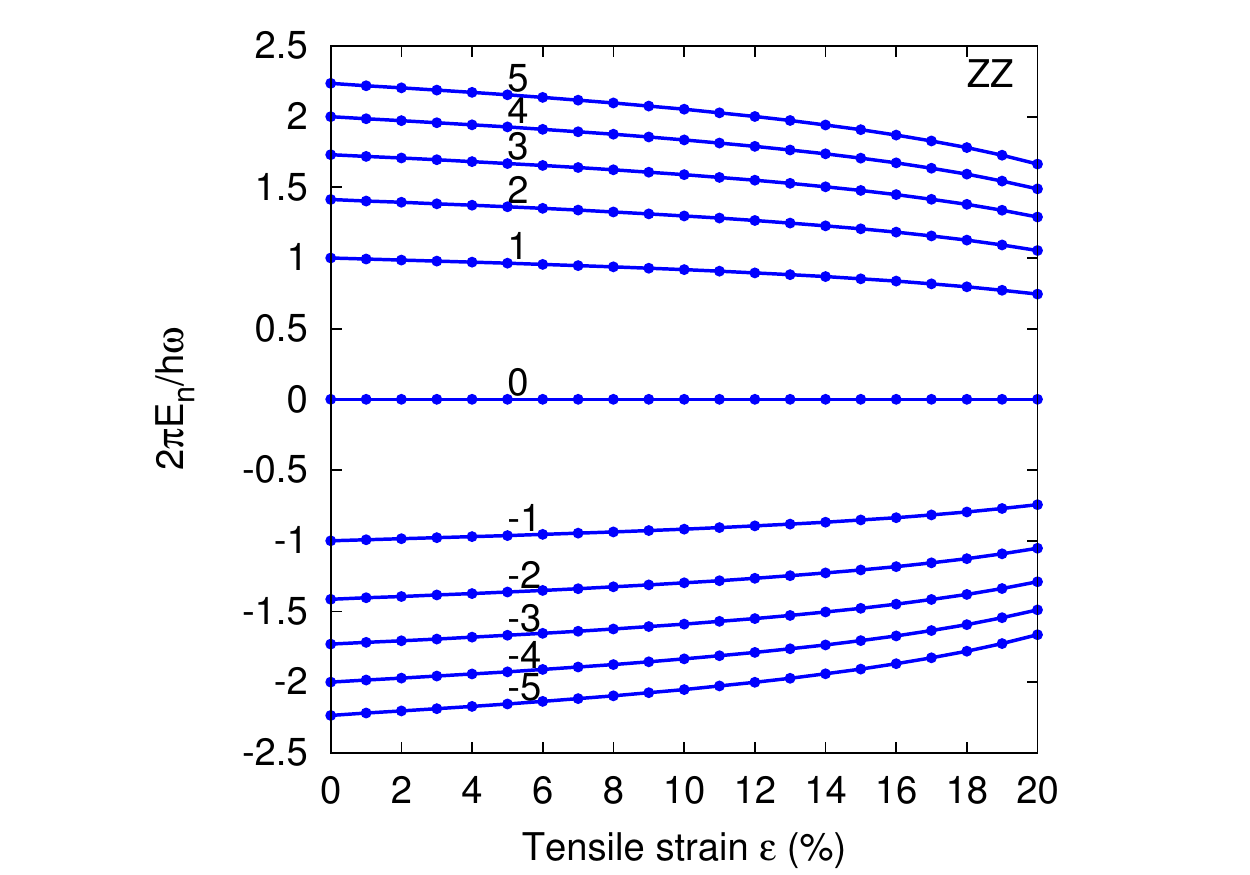}
\end{tabular}
\caption{{\small Evolution of the LLs spectra as a function of tensile strain along the (a) AC (red) and (b) ZZ direction (blue). The label in each curve corresponds to the Landau level ($n$).}}
\label{evo}
\end{figure*}
The contraction of the LLs spectra as a function of the tensile strain is shown in the Figure \ref{evo} for both strain directions (AC and ZZ). As we can see, the contraction of the LLs spectra for the ZZ deformation is larger than for the AC deformation. The LLs spectra does not present lifting of the two-fold valley degeneracy since our prediction is based in the low energy and low magnetic field regime \cite{Dietl, gail}. 

\begin{figure*}[t!]
\centering
\begin{tabular}{cc}
(a) \qquad \qquad \qquad \qquad \qquad \qquad \qquad \qquad& (b) \qquad \qquad \qquad \qquad \qquad \qquad \qquad \qquad\\
\includegraphics[trim = 14mm 0mm 18mm 0mm, scale= 0.79, clip]{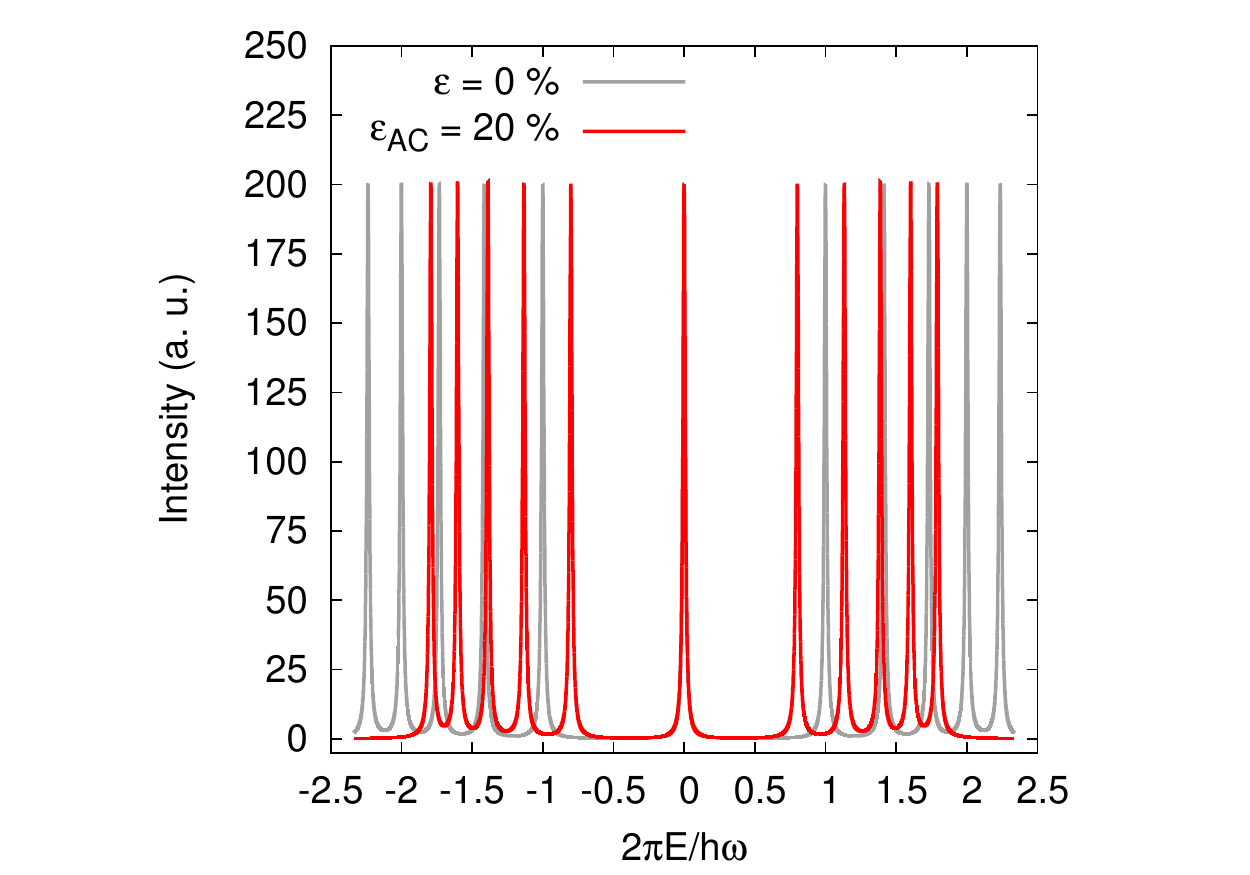} &
\includegraphics[trim = 14mm 0mm 18mm 0mm, scale= 0.79, clip]{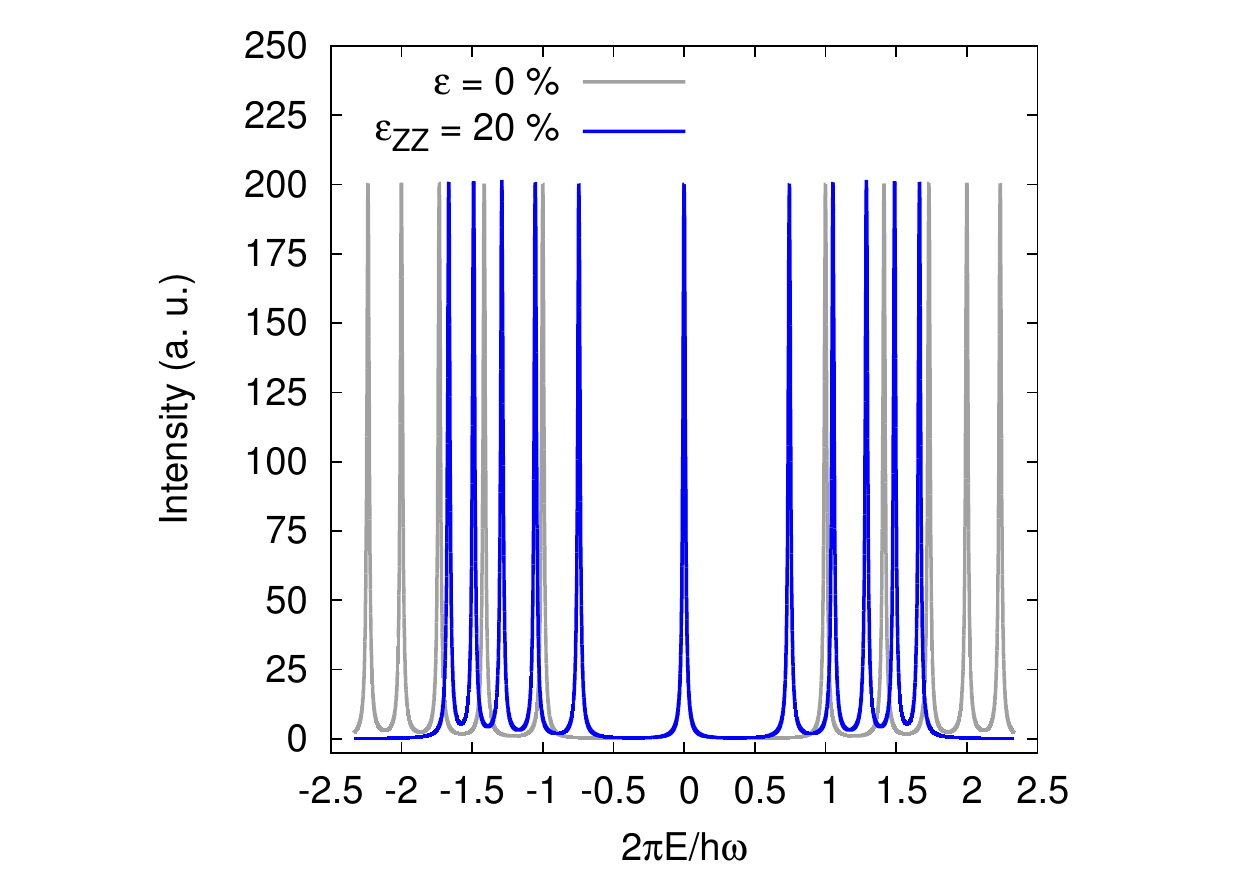}
\end{tabular}
\caption{{\small Calculated LLs density of states (DOS) for (a) pristine graphene (gray) and uniaxially strained along the AC direction (red) for deformation of 20\% ($\xi_{AC} = 0.801$) and (b) pristine graphene (gray) and uniaxially strained along the ZZ direction (blue) for deformation of 20\% ($\xi_{ZZ} = 0.745$).}}
\label{DOS_LLs}
\end{figure*}

In order to show the effect of strain on the LLs Density of State (DOS) under a uniform magnetic field, we have calculated the DOS for both strain directions  with a deformation of 20\%, and comparing them with the corresponding to pristine graphene under the same magnetic field. The DOS for strained graphene along the AC and ZZ directions are shown in the Figures \ref{DOS_LLs}(a) and \ref{DOS_LLs}(b), respectively. In these plots, we observe that the contraction of the LLs spectrum is larger along the ZZ than for AC direction, because $\xi_{ZZ} = 0.745$ and $\xi_{AC} = 0.801$ for $\epsilon = 20$\%. The DOS contraction with respect to equilibrium spectrum is a result of the LLs spacing reduction due to the modulation of the Fermi velocity by the strain. Thus, the present results show clearly that the LLs spectra in graphene can be modulated via uniaxial strain.

\section{Summary and final remarks}

We have proposed a geometrical approach to consider the deformation of the Dirac cones in strained graphene introducing a renormalized linear momentum in the effective Dirac Hamiltonian for massless fermions. We found an analytical expression for the energy spectrum of the LLs, which is a function of the Dirac cones deformation. In particular, we found that uniaxial deformation in graphene induces contraction of the LLs spectra for both AC and ZZ strain directions. The present findings help to set the Landau levels spectroscopy in strained graphene, which could be used for measure quantities of interest as the anisotropic Fermi velocity, Hall resistance and related electronic properties \cite{ImgLLs}. The present model offers a simple way to relate electronic, vibrational and transport properties with the geometry of the anisotropic Dirac cones. Finally, it is important to note that the present approach can be applied to analyze other two-dimensional materials beyond graphene presenting Dirac cones \cite{DC3D}.

\section*{Acknowledgments}

The authors Y. B-O. and M.E. C-Q. gratefully acknowledge a graduate scholarship from Consejo Nacional de Ciencia y Tecnolog\'ia of M\'exico (Conacyt-M\'exico). This research was supported by Conacyt-M\'exico under Grant No. 83604. Computational resources were provided by “Cluster H\'ibrido de Superc\'omputo - Xiuhcoatl” at Cinvestav. The authors thanks to A.G.S.O. de Montellano, S. Fern\'andez-Sabido and C.M. Ramos-Castillo, for a critical reading of the manuscript.





\end{document}